# A New Approach for Improvement Security against DoS Attacks in Vehicular Ad-hoc Network

Reza Fotohi
Young Researchers and Elite Club
Germi Branch, Islamic Azad University
Germi, Iran
Fotohi.Reza@Gmail.com

Yaser Ebazadeh
Department Of Computer Engineering
Germi Branch, Islamic Azad University
Germi, Iran

Mohammad Seyyar Geshlag
Department Of Computer Engineering
Shabestar Branch, Islamic Azad University
Shabestar, Iran

*Abstract*— Vehicular Ad-Hoc Networks (VANET) are a proper subset of mobile wireless networks, where nodes are revulsive, the vehicles are armed with special electronic devices on the motherboard OBU (On Board Unit) which enables them to trasmit and receive messages from other vehicles in the VANET. Furthermore the communication between the vehicles, the VANET interface is donated by the contact points with road infrastructure. VANET is a subgroup of MANETs. Unlike the MANETs nodes, VANET nodes are moving very fast. Impound a permanent route for the dissemination of emergency messages and alerts from a danger zone is a very challenging task. Therefore, routing plays a significant duty in VANETs. decreasing network overhead, avoiding network congestion, increasing traffic congestion and packet delivery ratio are the most important issues associated with routing in VANETs. In addition, VANET network is subject to various security attacks. In base VANET systems, an algorithm is used to dicover attacks at the time of confirmation in which overhead delay occurs. This paper proposes (P-Secure) approach which is used for the detection of DoS attacks before the confirmation time. This reduces the overhead delays for processing and increasing the security in VANETs. Simulation results show that the P-Secure approach, is more efficient than OBUmodelVaNET approach in terms of PDR, e2e_delay, throughput and drop packet rate.

## I. INTRODUCTION

VANETs is particular, MANETs by which where vehicles and fixed location at the roadside can keep in touch speak with each. It can self-structure, spread comfortably and cost low with open structures. VANET can pleasure an increasingly significant role in multiple regions: when an event an episode occurs, it sends the procedure message speed to other cars besides the procedure regions that actually help to prevent crashes again; cars catch vehicle velocity, density status of several roads, and they can they are able to take actions ahead of time which facilitates traffic congestion; also, cars can surf the internet via the fixed stations at the roadside. As a result of more and more apps, VANET has turned into an into the focus of research institutes and scientists worldwide [1-5]. Each year there are more and more traffic jams on the roads. This is large due to every year there are more and more cars on the streets so that in 2013 there will be1410 a million vehicle in the world. It is feasible to find conditions where communications between cars can help to prevent accidents. In this approach, cars can help with these questions every week. This can prevent great wastes of time, money and of oil reserves, in addition, governments spend lots of money and destroy the landscape when creating more roads because existing roads do not support the generated traffic. The resultant, a restructuring of traffic can prevent some of the aforementioned dilemmas. There are many points to consider in any wireless networks in overall. The Figure following gives an illustration of VANETs.

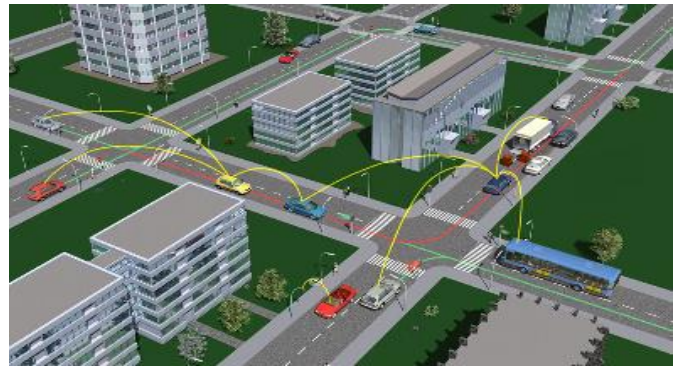

Fig. 1. An example of VANETs senario.

In VANET's safety is a significant issue that needs to be taken into account when any wireless network is designed. VANETs have weaknesses to various kinds of DoS attacks [6], Blackhole, gray hole and wormhole portion of the region of the safety problems existent in this kind of networks. Nevertheless, an another approach is presented in this paper. The main purpose of this work is to define a proposed approach, a scalable free system for VANETs where users can cooperate via their mobile technology and obtain updated information of interest about the traffic and attacks area in order to choose the best-refreshed path to their goals. in this paper, we proposed P-Secure approach detection algorithm is that attacks, for detecting DOS attacks used to commit time. This decreases the overhead for processing and securing the VANET is delayed. This approach has better special depending criteria removal rate, throughput, PDR and latency to develop e2e_delay, Therefore, this work proposes a self-managed VANET without any infrastructure, which will serve as an introduction to a more complex VANET, all this with better levels of security. In this paper, we implement the proposed P- Secure approach as the solution in NS-2 simulator to test its performance it. In the second section of paper, related works are presented. Afterward, section 3, Dos Attack in Vehicular Ad-hoc Network (VANETS), section 4, The proposed method, Section 5 Experimental Data and Analysis, Finally, in section 6 conclusion.

## II. RELATED WORK

Security in the network is of specific problems due to man lives are permanently at the condition as in traditional networks the major security concerns include confidentiality, integrity, and availability none of which involves primely with life security. Essential information can't be either changed nor deleted by a malicious node. Yet, security in the network also contains the ability to specify the driver responsibility while maintaining driver privacy. Information about the car and their drivers within must be swapped securely and more importantly, timely in that the delay of message exchange may cause catastrophic consequences such as the collision of vehicles. The spread of a general security model for the network is very challenging in practice.

With its dynamic characteristics and high mobility, the usage of wireless technology also makes VANET vulnerable to DoS attacks that exploit the open and broadcast characteristics of wireless networks. [7] Cryptographic attacks in VANET are classified in the next section. Further common networks security problem, unique security challenges arise because of the unique characteristics of VANET such as high mobility, dynamic topology, short connection duration and frequent disconnections. These novel characteristics bring safety concerns such as trust group formation, location detection, and security as well as certificate management. Corresponding preview work will be given in following sections based on the characteristics of the protection issue in similar work. The clustering system has been well thought in wireless technology in recent years [8]. Nevertheless, considering the natures of VANETs, such as sufficient energy, high speed, the clustering methods proposed for conventional wireless networks are not proper for VANETs. Hence, the clustering approach for VANETs should be designed exactly. The lowest ID clustering algorithm [9] is one of the easiest methods to cluster mobile nodes for VANETs. Using this method, all of the nodes broadcast becomes stages in which the node IDs are encapsulated. Further, these nodes IDs are assigned uniquely. In [10], authors proposed a clustering approach using affinity propagation for VANETs. Affinity propagation is first proposed to solve data clustering problem and it is show that this algorithm can generate clusters more efficiently compared with traditional solutions. The node which has the lowest ID in its neighborhood is selected as the cluster head node, and other nodes are selected as the cluster member nodes. The lowest ID algorithm proposed a basic idea to cluster mobile nodes. First, we need to define a metric to model the property of wireless nodes; and then we can use the metric to group nodes based on some rules. The follow- ing clustering schemes are all based on this basic idea. The difference of various clustering scheme is the metrics used for modeling. Density Based Clustering (DBC) algorithm is proposed in [12]. Using DBC, connectivity level, link quality and traffic conditions are taken into account completely to cluster vehicle nodes. The mobile network is divided into the dense part and sparse part. A node which has links more than a predefined value is considered as in the dense part; otherwise, it is in the sparse part. During the clustering process, link quality is estimated to make a re-clustering decision. According to the experiment results, the cluster head change ratio is less than the lowest ID algorithm [13]. So, there are some works devoted to design and develop specific simulators of VANET. Groove Net [14] is a hybrid simulator which enables communication between simulated vehicles, real vehicles and between real and simulated vehicles, and it models inter-vehicular communication within a real street map-based topology which is based TIGER map data. MOVE and Translated [15] can rapidly generate realistic movement model which can be directly used in network simulators such as NS2, SUMO. VG Sim [16] combines movement model of vehicles and network simulation and transforms vehicular moves and applications to events for further processing of network simulators. NCTUns [17][18] Different kinds of attacks have been analyzed in MANET and their effect on the network. Attack such as grayhole, where the attacker node behaves maliciously for the time until the packets are dropped and then switch to their normal behavior [19]. MANET's routing protocols are also being exploited by the attackers in the form of flooding attack, which is done by the attacker either by using RREQ2 or data flooding [20]. Design and presentation of different security obstacles and attacks in mobile ad hoc networks as well as finding appropriate solutions to them is a challenging research area for researchers. Black hole attack is one of the famous related attacks. In [11], the idea of affinity propagation is used to cluster vehicle nodes in a distributed manner. The vehicle nodes exchange messages with their neighbor nodes to transmit availability and responsibility and make the decision based on the availability and responsibility values for constructing clusters. The simulation results demonstrate that the performance of the clustering scheme using affinity propagation is better than MOBIC in terms of stability.

## III. DOS ATTACK IN VEHICULAR AD-HOC NETWORKS

In a VANETs, usually the attacker attacks the communication medium to cause the channel jam or to create count obstacles for the nodes from approachability the network. The essential purpose is to forbid the authentic nodes from approachability the network services and from using the network sources. The attack would result in failure of the nodes and VANET sources. Finally, the VANETs are no longer available to legitimate nodes. In VANETs, DoS should'nt be permissible to happen, where seamless life critical information must reach its intended destination securely and timely. In short, there are 3 routes the attackers could get DoS attacks, scilicet communication channel, network overloading, and dropping the packets [8]. In calculating, a DoS attack is an attempt to make a system or VANET source unavailable to its intended users, like to temporarily cut off or suspend tasks of a host connected to the network. A DDoS is where the attack source is more than one, often hundreds of, unique IP. It is similar to a set of people crowding the entry door and not letting legitimate parties enter into the store, disrupting ordinary tasks. sinful perpetrators of DoS attacks mostly target services hosted on high-profile web servers such as banks, credit card payment gateways; but motives of revenge, blackmail or activism can be behind other attacks. in Dos Attackers diffuse false information to affect the behavior of other drivers In Figure 2 is shown an example of DOS attacks.

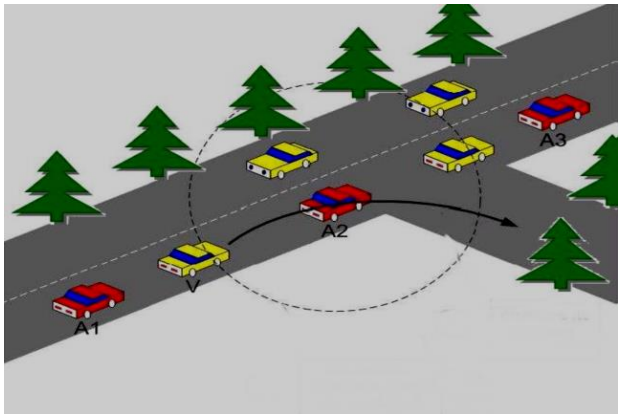

Fig. 2. An example of the problems in the Dos attacks in VANET

In Figure 2, A2 sends a false message with traffic info, and V changes its route and frees the road, in conclusion, Attackers tracks vehicles to obtain those drivers' private information ,and routing disrupts DOS efforts of to influence the sentences are as follows:

A. Types of outbursts network include TCP, UDP, and ICMP that disrupt legitimate traffic site

B. Trying to disconnect the machine and thus not being able to use their service

C. Trying to prevent a particular individual from accessing a service.

D. Trying to sabotage the service to a specific system or person

E. Trying to disrupt the hair routing network

## IV. THE PROPOSED METHOD (DETAILS)

The essential aim of the VANETs is to supply security and welfare for the travelers. To achieve this object, a specific electronic device is embedded in any vehicle that makes it possible to establish communication between the passengers. Such a network must be implemented without client-server network limits of communication structures. Each vehicle armed with a VANET machinery is a node in a mobile wireless network and is able to receive and send others' messages via the wireless network. Traffic alerts, road signs and traffic observation for a moment that can be transmitted through such a network, provide the necessary tools to make decisions about the best path for the driver. In this paper we proposed a approach in a vehicle network improves road safety, transport efficiency, but also reduces the impact of transport on the environment; all three of these applications are not perfectly perpendicular to each other. For example, reducing the number of accidents, in turn, can reduce the traffic congestion and this can lead to a reduction of the environmental impact.

### A. The proposed approach

At first, we must have a system model to express, the proposed method based on which we know the position of the vehicles and roadside equipment as well as antennas etc. We introduce our proposed method with the name P -Secure Protocol that has been named as P-SP in the Figure.

### B. The proposed system model

Figure (3) indicates a road that has radio transmissions and vehicles that have onboard radio. RSRU (roadside radio unit) decides to depend on its transmission range and set up in the area where the vehicles can form a network. In fact, we consider it as a threshold. The vehicles can send messages to RSRU through the proposed P-Secure Protocol mechanism. In this way, detection of the exact position of the vehicle that has sent the message is conducted. After discovering the situation, the vehicle's data are saved in some RSRU. Any vehicle with OBU and anti-tampering sensors is (Tamper PROOF). These devices have the responsibility of storing the accurate information about the vehicle like speed, location and more. The position of the vehicles is obtained through the frequency and speed of vehicles and the use of OBU. Vehicles can use the P-Secure Protocol mechanism to request from RSRU. In fact, RSRU conducts the approval of vehicles and maintains the database. Location, time, and etc. along with the data package are provided to RSRU. Traffic devices use the requests using another database and provide the service responses only to the radio transmitter which has been approved, therefore causes the reduction of DOS attacks which could be created due to Flooding.

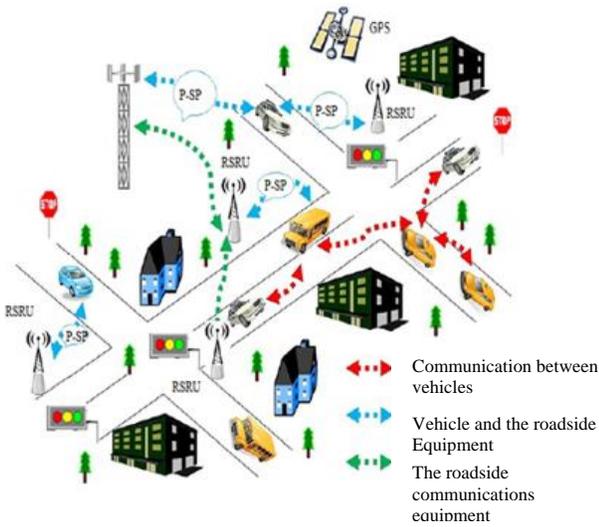

Fig. 3. The proposed system model (RSRU range of vehicles and a process in which the request is sent)

## C. The phase 1 of the P-Secure approach

Proposed P-Secure, discovers the vehicle location and information packets that the vehicles have sent and if the packet was not related to the attack, the vehicle is not detected. To conduct the algorithm, it is required to consider a number of variables:

**Definition 1:** We consider a maximum packet capacity and display it with M where we consider the value 20 for our work.

**Definition 2:** p parameter is the number of packets sent to RSRU per second.

**Definition 3:** α is the coefficient which is determined by the characteristics of the road.

**Definition 4:** V_m is the maximum speed of the vehicle.

**Definition 5:** V is the speed of the vehicle.

In addition to the maximum speed, a minimum speed is also required that it demonstrated by low. To obtain the number of packets that are sent by the RSRU vehicles per second (P) the Equation (1) is used:

$$P = \alpha * \left|\frac{V - vm}{2}\right| \tag{1}$$

Since the number of packets and the maximum speed is higher than the nod speed, the position of the vehicle is changing rapidly. We consider this as an attack and if the speed is too low, the vehicle's position will not change much and we consider this as an attack. For the attack detection algorithm, we act as the following steps:

1. First we get the required information from the vehicles
2. We set the threshold for RSRU
3. The confirmation requests are sent to RSRUs by the vehicles.
4. A time stamp is expressed for each device.
5. If (the time stamp of the transmitter – time stamp of the receiver) is greater than the threshold value, the packet is discarded, otherwise the package is acceptable.
6. Have the value of M equal to 20 and find the p value for all packets less than or equal to 20.
7. If the p value is greater than or equal to 20 and the v value is greater than or equal to the maximum speed (v_m), the packet is then discarded and the packet is detected as an attack.
8. The next mode of attack is that: if the p-value is less than or equal to 20 and the v value is less than or equal to the minimum speed (low), then the packet is discarded and the packet is detected as an attack.

### D. phase 2 of the P-Secure approach

In phase 2 of the P-Secure approach, we detected attacks the packet capacity and/or their speed has been more or less than the minimum whereas the probability of DOS attack also exists in the delivery of any vehicle. To detect these attacks, we improve the proposed algorithm and add the following steps to it:

TABLE I. REQUAREED PARAMETERS

| | |
|---|---|
| p | The number of packets per second |
| M | The maximum capacity packet |
| α | Coefficient |
| $v_m$ | Maximum speed |
| V | Speed |
| low | Minimum speed |

1. The confirmed vehicles are buffered at previous stages in the RSRU
2. Any vehicle wishing to enter the VANET network must submit its application RSRU
3. RSRU updates its' counter for that vehicle.
4. Allocating time slot to the devices is approved by RSRU
5. RSRU counts the number of steps of the vehicles; if the number of steps is equal to the number of counting times, the counter then updates the vehicle's next step and declares the vehicle as valid.
6. The new requests are assessed.
7. If the vehicle has a new request in the same time slot and/or the request number of this vehicle is more than the other vehicles, it is identified as malicious.
8. Indicates the number of the damaging vehicle as the malicious.

P-Secure approach discovers DoS attack before the confirmation stage. Location, time stamp, speed and etc. of the vehicle are considered to find out whether it is within the radar range or not. The proposed algorithm is also used in detecting the false alerts. If the count of packets and the MAX speed are above the node's speed, it is considered as an attack on the position of the vehicle is changing quickly. Likely, if speed is too low, the vehicle's position will not change much and this is also considered as an attack. After completing the process of valid vehicles, they are stored in the RSRU database. The phase 2 of the approach is to confirm the new request that wants to join the network. This algorithm compares the previous valid database with new requests and reduces false requests through allowing valid nodes. The proposed P-Secure Protocol algorithm reduces attacks by limiting the counter and also not allowing the fake vehicles by the attackers. The flowchart of the proposed method is fully displayed in Figure (4).

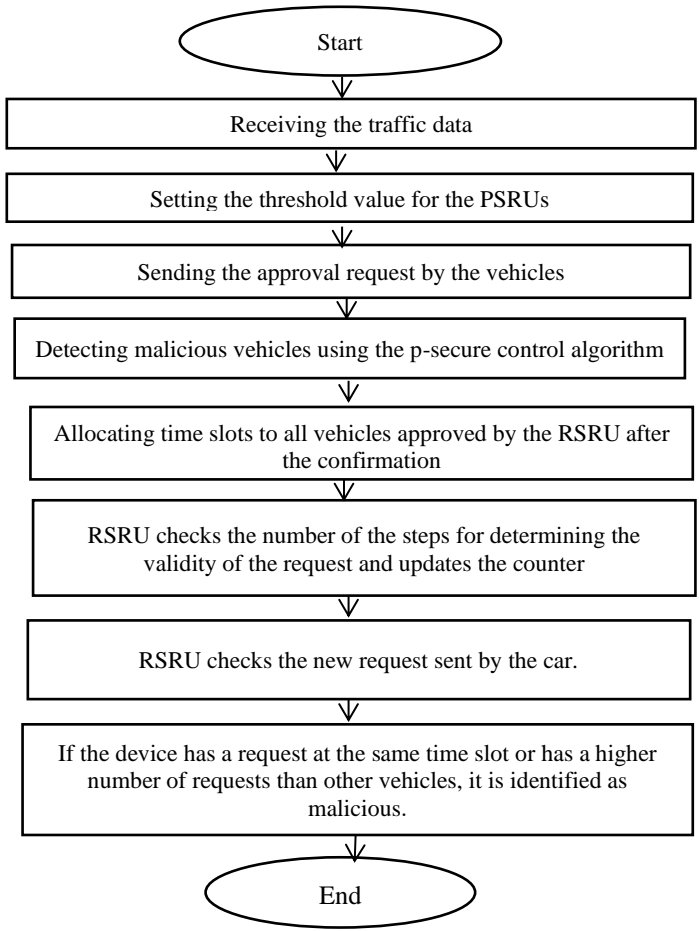

Fig. 4. Flowchart of the P-Secure approach

## V. SIMULATION RESULT AND ANALYSIS

This part contains evaluation of P-Secure approach compared with OBU model VANET approach. With the help of the NS-2 we were able to prove, ns is simulator project, start 1989 as a different of real. We run two senario, one P-Secure approach, and OBUmodelVaNET approach, we have repeated the experiments by changing the several times. To, 200, 400, 800, 1000, and 1200. The simulation parameter are shown in table 2 the parametess used to implementation the performance are given follows.

TABLE II. SIMULATION PARAMETERS

| Simulator | NS2.34 |
|---|---|
| Area | 1000m X1000m |
| Density | 150-20 |
| Transmission Range | 250m |
| Antenna | Omni Antenna |
| Simulation duration | 200,400,600,800,1000,1200 |
| MAC Layer | 802_11 |
| Traffic Type | CBR (UDP) |
| Buffer Size | 150 Packet |
| Node placement | Random |
| Simulation methods | *OBUmodelVaNET* AND *P-Secure approach* |

As we saw in the previous section, DOS attack causes weakness in making the network unsecure and weakening the ordinary approach of the network, in this article we deal with providing a solution for security improvement in car networks against denial-of-service attacks which improves the standard end-to-end delay, packet delivery ratio, the number of drop packets and throughput.

### A. PDR

PDR is define as shown in $Equation$ (2).

$$PDR = \frac{\text{Number of sent Packets}}{\text{Number of received Packets}} * 100 \quad (2)$$

Where PDR is the package delivery rate, SendPacketNo is the number of sent packages, and RecievePacketNo denotes the number of received packages.

### B. Packet Drop Rate

It is the measure of the number packets dropped due to malicious nod**e** (DoS attack). Thus, we can define $Thro$ as shown in $Equation$ (3).

$$Drop = \frac{\text{Send Packet} - \text{Received Packet}}{\text{Send Packet}} \quad (3)$$

### C. End to End Delay

End-to-end delay refers to the time taken for a packet to be transmitted Around the Network from source to destination.

### D. Throughput

Throughput is the number of data packets transmited from source node to destination node [21]. Thus, we can define $Thro$ as shown in $Equation$ (4).

$$Thro = \frac{Request}{Time} \qquad (4)$$

Where **Thro variable** is the throughput, **requests** is the number of requests that are accomplished by the system, and **time** shows the total time of system observation.

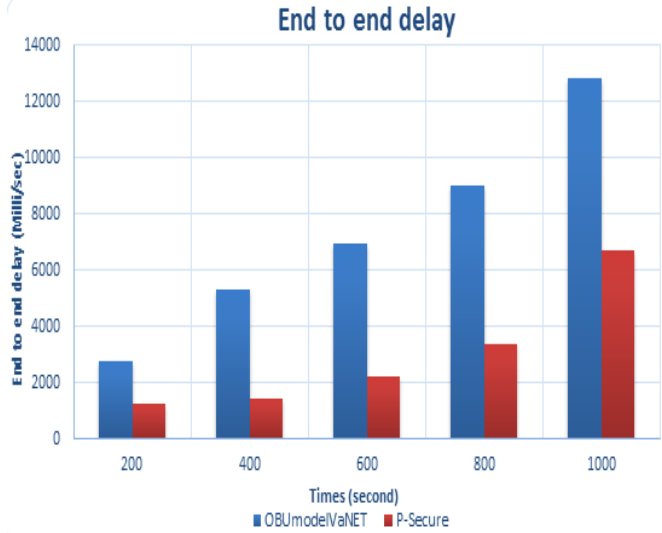

Fig. 5. E2E_delay vs time

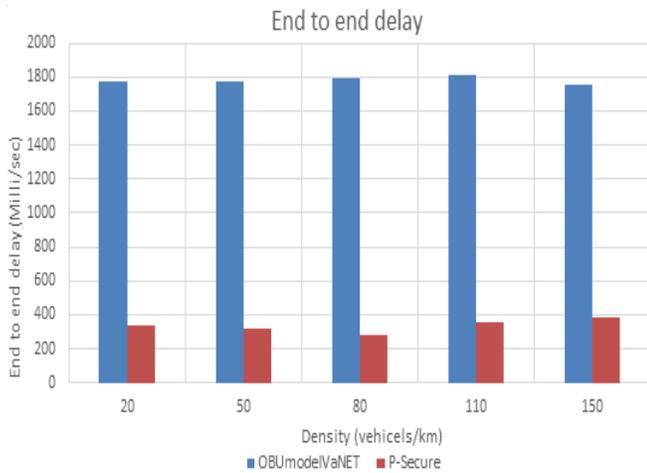

Fig. 6. End-to-end delay vs density

Fig (5) and (6), shows e2e_delay against the time and dencity. P-Secure approach is significantly lower compared to OBUmodelVaNET aproach. The reason is that using the proposed algorithms in the P-Secure approach, the suspicious nodes are not allowed to be sent and the attack was also diagnosed quickly and the security messages are delivered with a very little delay to the vehicles that are at risk.

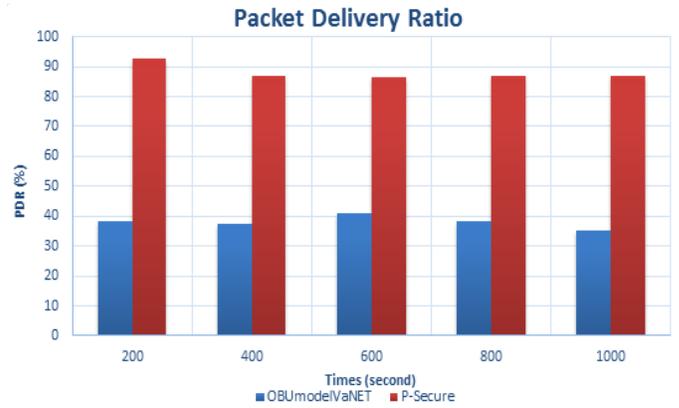

Fig. 7. PDR vs time

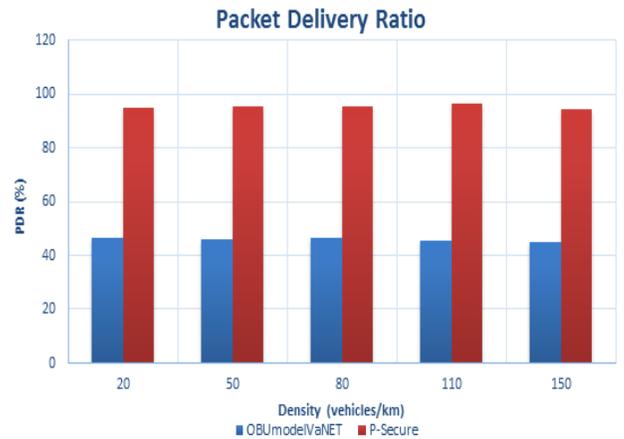

Fig. 8. Packet Delivery Ratio vs density

Fig (7) and (8), shows PDR against the time and dencity. From following graph, we can say that value of PDR is decreasing for OBUmodelVaNET under attack. When we vary pause time from 200 to 1000 as well PDR values for proposed approach is high. Thus, we can say that PDR of proposed method is improved than OBUmodelVaNET under attack which degraded due to DoS attack.

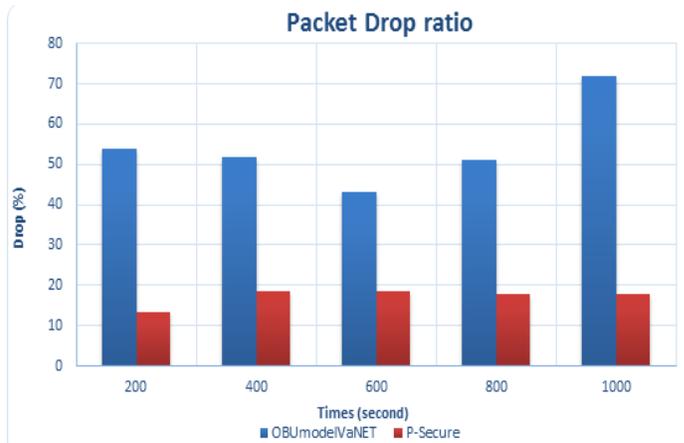

Fig. 9. Dropped Packet Rate vs time

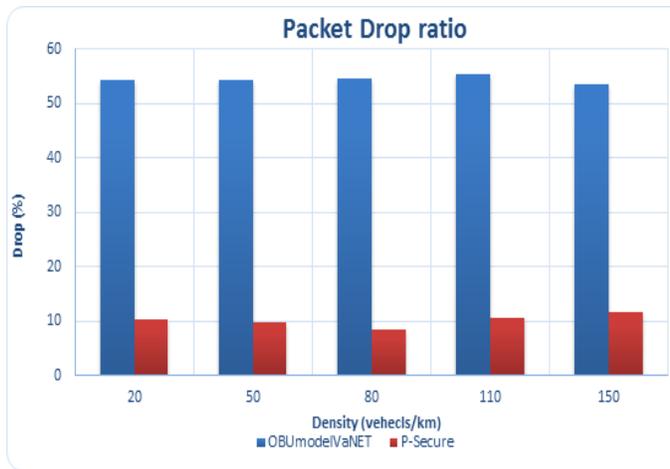

Fig. 10. Dropped Packet Ratio vs density

Fig (9) and (10), shows drop ratio against the time and dencity. It shows that, between the pause times 200 to 1000, the OBUmodelVaNET under attack had a high packet drop, while the packet drops of proposed approach in these times, has decreased.

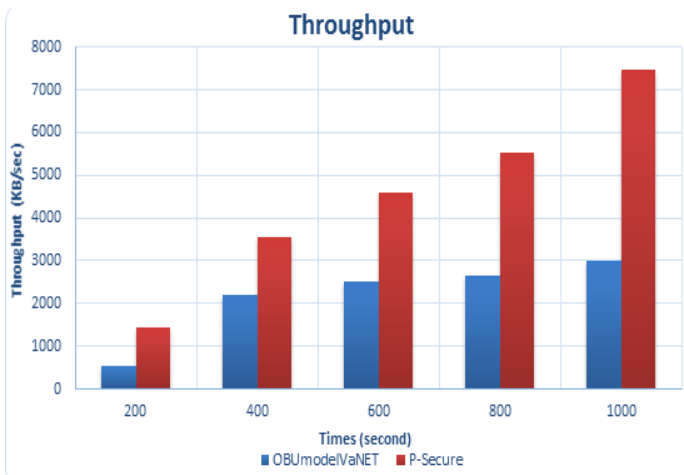

Fig. 11. Throughput vs time

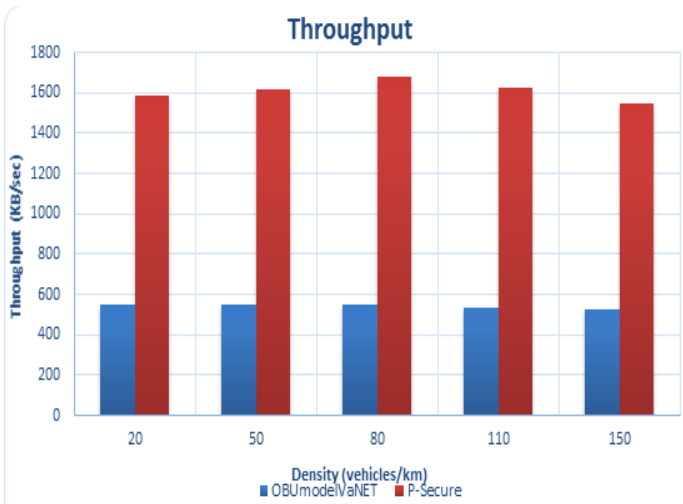

Fig. 12. Throughput vs density

Fig (11) and (12), shows throughput against the time and dencity. So we can claim that the P-Secure approach has better performance in the field of throughput rather than OBUmodelVaNET approach in MANET. This is due to the proposed mechanism that shown in fig 4.

## VI. CONCLUSION

VANETs are a kind of wireless networks which have been created to communication between the adjacent vehicles as well as adjacent vehicles with constant equipment which are usually roadside equipment. One of the main concerns in Vehicular ad hoc networks (VANETs) is the attacks and influence to the system and of course having safety from the key concerns for many road users. In this paper, we introduce the (P-SECURE PROTOCOL) attack detection algorithm that is applied for detecting DOS attacks before the confirmation time. In the proposed RSRU method depending on its transmission range and using the proposed algorithm decides on what is the secure message, and sets the area where the vehicles can form a network. Traffic devices use the requests using another database and provide the service responses only to the radio transmitters which are approved therefore lead to the reduction in DOS attacks. The P-Secure approach leads to the reduction in processing delays and improving safety in VANET. To demonstrate the performance of P-Secure approach using the NS-2 simulator, the proposed system is compared to OBUmodelVaNET approach. The simulation results shows that P-Secure approach outperform than OBUmodelVaNET approach in terms of e2e_delay, PDR, packet drop rate and throughput.